\documentclass{article}

\usepackage{PRIMEarxiv}
\usepackage[ruled,vlined]{algorithm2e}
\usepackage[utf8]{inputenc} 
\usepackage[T1]{fontenc}    
\usepackage{hyperref}       
\usepackage{url}            
\usepackage{booktabs}       
\usepackage{amsfonts}       
\usepackage{nicefrac}       
\usepackage{microtype}      
\usepackage{lipsum}
\usepackage{fancyhdr}       
\usepackage{graphicx}       
\graphicspath{{media/}}     
\usepackage{multirow}
\usepackage{amsmath}  
\usepackage{subcaption} 
\usepackage{bm}
\usepackage[ruled,vlined]{algorithm2e}
\usepackage{algorithmic}

\begin{document}
\onecolumn

\title{Functional Graph Contrastive Learning of Hyperscanning EEG   Reveals Emotional Contagion Evoked by Stereotype-Based Stressors}



\author{
  Jingyun Huang$^1$, Rachel C. Amey\,$^{2}$, Mengting Liu$^3$\thanks{Corresponding author}, Chad E. Forbes$^4$\\
  $^1$\textit{School of Computer Science and Engineering, Sun Yat-sen University, Guangzhou, China}\\
  $^{2}$Army Research Institute for the Behavioral and Social Sciences, Fort Belvoir, VA, USA\\
  $^3$\textit{School of Biomedical Engineering, Sun Yat-sen University, Shenzhen, China}\\
  $^4$\textit{Department of Psychology, Florida Atlantic University, Boca Raton, FL. 33431, USA}\\
}

\renewcommand*{\thefootnote}{*}


\maketitle

\begin{abstract}
     This study delves into the intricacies of emotional contagion and its impact on performance within dyadic interactions. Specifically, it focuses on the context of stereotype-based stress (SBS) during collaborative problem-solving tasks among female pairs. Through an exploration of emotional contagion, this study seeks to unveil its underlying mechanisms and effects. Leveraging EEG-based hyperscanning technology, we introduced an innovative approach known as the functional Graph Contrastive Learning (fGCL), which extracts subject-invariant representations of neural activity patterns from feedback trials. These representations are further subjected to analysis using the Dynamic Graph Classification (DGC) model, aimed at dissecting the process of emotional contagion along three independent temporal stages. The results underscore the substantial role of emotional contagion in shaping the trajectories of participants' performance during collaborative tasks in the presence of SBS conditions. Overall, our research contributes invaluable insights into the neural underpinnings of emotional contagion, thereby enriching our comprehension of the complexities underlying social interactions and emotional dynamics.
     
\end{abstract}


\section{Introduction}


Emotional contagion refers to the sharing of emotional states between individuals, and it has been observed in both animal and human models that the infectivity of negative emotions is much greater than that of positive emotions \cite{Goldenberg2020}. Negative emotional contagion has a powerful effect on our relationships - family, friends, teams, etc. - and can lead, for example, to depressive behavior in healthy people who live with depressed individuals. It is urgent to understand the mechanism of emotional contagion, especially negative emotional contagion. Emotional contagion has long been regarded as reflecting a mimicry-based process, for which mimicry of emotional expressions and its consequent feedback function are assumed and can be evoked by higher-order social processes or by a simple emotion-to-action response as well as the primary mimicry-based process \cite{10.3389/fpsyg.2022.849499}. At present, the emotional contagion models mostly adopt behavioral analysis and questionnaires, which are often affected by subjects' subjective factors. They have mainly focused on behavioral experiment such as analysing people's posts containing emotional information to extract affective evidence \cite{doi:10.1073/pnas.1320040111}, using the Positive And Negative Affective Schedule (PANAS) scale to measure positive and negative emotions as a quantitive research \cite{8035125} and the mathematical simulation model of emotional contagion in crowd evacuation \cite{8359363}. 

Although behavioral analysis and questionaires can provide valuable insights into emotional contagion, they have limitations in terms of capturing the neural mechanisms, timing, and subtleties of this phenomenon. To overcome these limitations, researchers have turned to EEG-based hyperscanning, a technology that records electroencephalographic (EEG) data from multiple participants simultaneously. This approach complements traditional behavioral analysis and questionnaires by providing a more direct and precise real-time examination of the underlying brain activities associated with emotional contagion. EEG-based hyperscanning technology has proven effective in capturing brain states during affective communication. For instance, when an individual experiences specific emotions like sadness, joy, or fear, their brain activity may influence the brain activity of others they interact with, thus bearing implications for emotional contagion \cite{Liu2018}.

However, the significant inter-subject variability of emotion-related EEG signals poses a great challenge for cross-individual emotional representation extraction \cite{Shen_2022}. In most cases, the accuracy of the intra-subject emotion classification is higher than that of inter-subject classification with the same classifier \cite{8882370, Suhaimi_article}. This limitation in the generalization of emotion classifiers may be attributed to individual differences in EEG-based emotional representations, influenced by factors such as personality, dispositional affect, and genotype \cite{HAMANN2004233}. Furthermore, individual variance in patterns of brain connectivity reveals that the inter-subject contrast plays a significant role in cognitive analysis \cite{Finn_Shen_Scheinost_Rosenberg_Huang_Chun_Papademetris_Constable_2015}. These previous findings suggest that EEG emotional representation should not be extracted without considering the individual difference.

In this study, we investigated whether women who experienced social identity threat could transmit their stress to women who were not under threat, using a process known as Stereotype-based Stress (SBS) contagion and examined how this collective stress affected women in a dyadic performance context. As previous research revealed, SBS contexts typically engender a variety of behavioral and physiological SBS responses including sustained vACC activation, unique neural network configurations and enhanced connectivity between regions integral for emotion (dACC, vACC, and mPFC) and saliency networks (IPL, insula, and STS) , this could collectively provide evidence for increased emotional processing and the awareness of negatively arousing or stressful information \cite{Forbes_Liu_article, Amey_article, Liu2020.05.26.117499, Liu2020.05.30.125476}. We sought to understand if threatened women can transmit their stress to otherwise non-threatened partners, does it hurt or benefit the woman directly under threat, and, to what extent can this come at a cost to their otherwise non-threatened partners? To this end, we designed an experimental paradigm of emotional contagion by using discussion and learning scenarios, and try to understand the physiological mechanism of emotional contagion by using EEG-based hyperscanning technology combined with a data-driven approach --- functional Graph Contrastive Learning (fGCL), which extracts the subject-invariant emotional representations while preserving Functional Connectivity (FC) information, with a downstream analysis to infer and explore its process from outcomes of emotional contagion. The formulation of fGCL is grounded on the assumption that the neural activities of the subjects are in a similar state when they receive the same segment of emotional stimuli (i.e., the displayed CORRECT or WRONG responses on the screen). Based on this fundamental idea, we aim to learn subject-invariant representations of EEG signals in the embedding space underlying similar mental activities. Specifically, fGCL mainly consists of two components, i.e., the spectral-based graph convolutional network (spectral GCN) encoder, and a two-layer multi-layer perceptron (MLP). It maximizes the similarities of the representations in response to identical emotional stimuli while minimizing the similarities between signals corresponding to different stimuli. In the downstream analysis, a classifier --- Dynamic Graph Classification (DGC) utilize the trained encoder extracted graph embeddings as input to identify brain state to emotional stimuli after emotional contagion within dyads. Since the representations are extracted based on semantically meaningful settings, they are expected to be informative and generalizable in the downstream analysis.   

The method presents three essential characteristics:
\begin{itemize}

 \item The presented model can extract EEG signal representations with inter-individual commonality and remove the individual differences, which more effectively summarize the internal neural activity pattern. 
 
 \item A deep learning model that is more effective than traditional statistical analysis and behavioral analysis methods to investigate the emotional contagion mechanism
 
 
 \item The graph data structure based analysis is more aligned with functional brain network structure and thus yield more intuitive and effective results
 
\end{itemize}


\section{Method}
\label{sec:headings}


\subsection{Graph Construction and Analysis Procedure}
As individual difference exists in inter-subject functional connectivity, a statistical dependency quantifies the connection strengths between brain region of interests (ROIs) \cite{Mueller2013}, our aim is to preserve the FC information for for more effective emotional analysis. To this end, we adopt graphs, which are naturally suitable for modeling brain topology. In this approach, we project ROIs onto the nodes of a graph and the weighted edges connect nodes, this overcomes the limitation of traditional 2D grid-like structure, where models might fail to explore and exploit the complex FC \cite{demir2021eeggnn}. 

Given a dataset $\{(\mathcal{G}_i^j, y_i)\}^N_{i=1}$ with N individuals, where $y_i = \{0, 1\}$ denotes the label of $i$-th graph, $\mathcal{G}_i^j = \{\mathcal{V}_i^j, \mathcal{E}_i^j\}$ is the $j$-th view in augmented views (see section \ref{sec:imp_de}) where Pearson correlations for each ROI is calculated as node features $X_i^j \in  \mathcal{R}^{ROIs \times ROIs}$ where $x_n = X[n, :]^T$ is the ROIs (the number of ROI in a graph) dimensional node feature of node $v_n \in \mathcal{V}$ containing FC information and Partial correlations between ROIs is used as edge features $\mathcal{H}_i^j \in \mathcal{R}^{ROIs \times ROIs}$ where $h_{n n'} = \mathcal{H}[n, n']^T$ is the edge feature of edge $e_{nn'} \in \mathcal{E}$.

\begin{figure}
  \centering
  \includegraphics[scale=0.5]{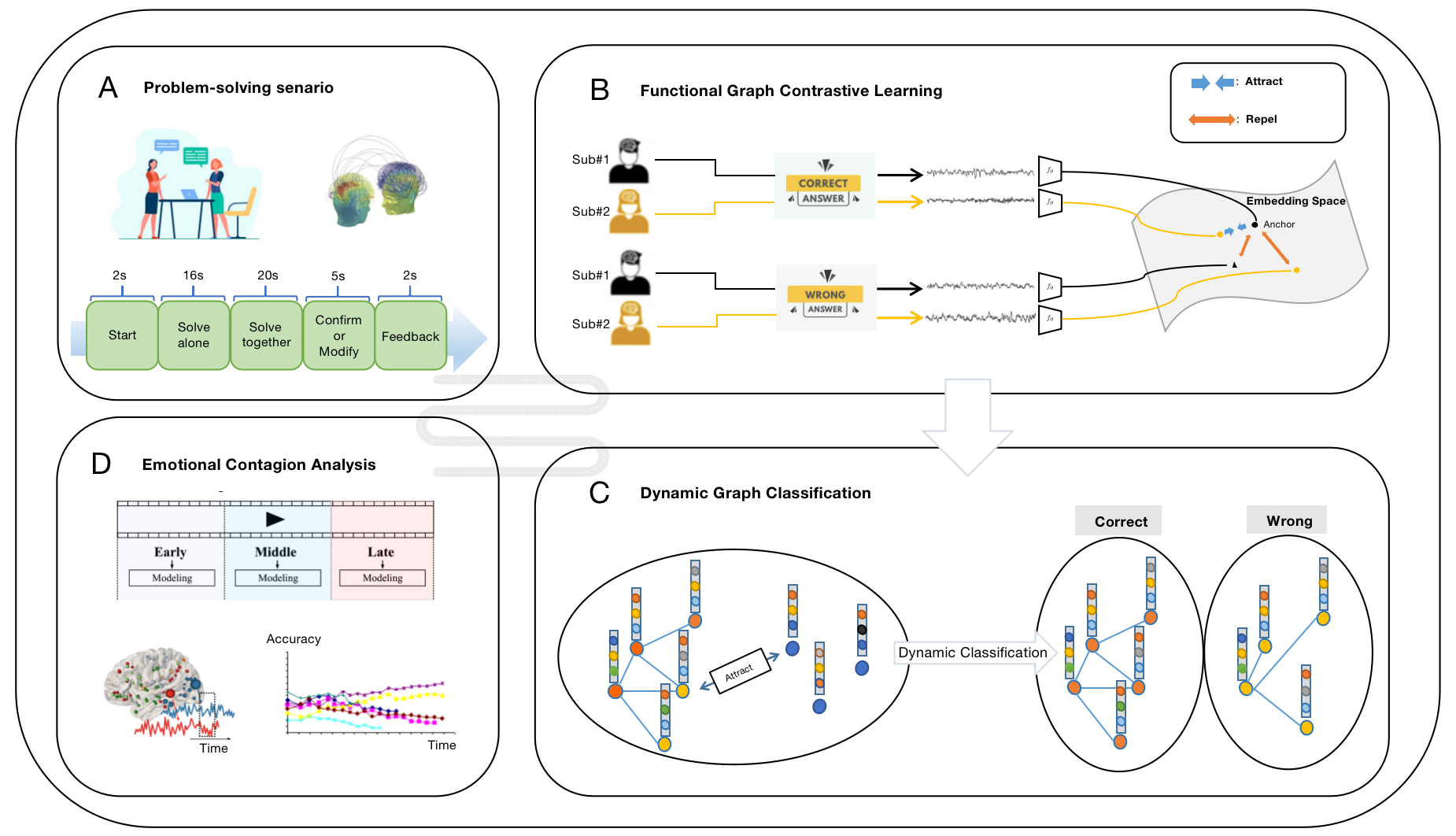}
  \caption{The workflow of our approach. (A) Experiment Setting. Subjects in one dyad answer mathematical questions while recording their EEG signals simultaneously. (B) Given a minibatch of graph data representing subjects in the same dyad, the fGCL encoder is able to extract subject-invariant graph embeddings. (C) The splitted graph embeddings are then fed into DGC classifier to produce classification evaluation results at these three stages respectively. (D) Donwstram analysis. Data are divided into early stage, middle stage and late stage according to the senario of tasks to perform further emotional contagion analysis.}
  \label{fig:workflow}
\end{figure}

As illustrated in Figure \ref{fig:workflow}, the functional Graph Contrastive Learning (fGCL) encoder learns to extract embedding of each graph as subject-invariant representations (i.e., maximize the representational similarity of EEG signals belonging to the similar scenario, and minimize that of others), by leveraging this encoder, we construct a population graph where a collection of subject-invariant embeddings of EEG signals as nodes. Given this, the Dynamic Graph Classification (DGC) classifier is trained to perform classification task on three stages (early, middle and late). The results can be further utilized for emotional contagion analysis (see Secction \ref{se:Contagion_Stage_Analysis}).

\begin{figure}
  \centering
  \includegraphics[scale=0.6]{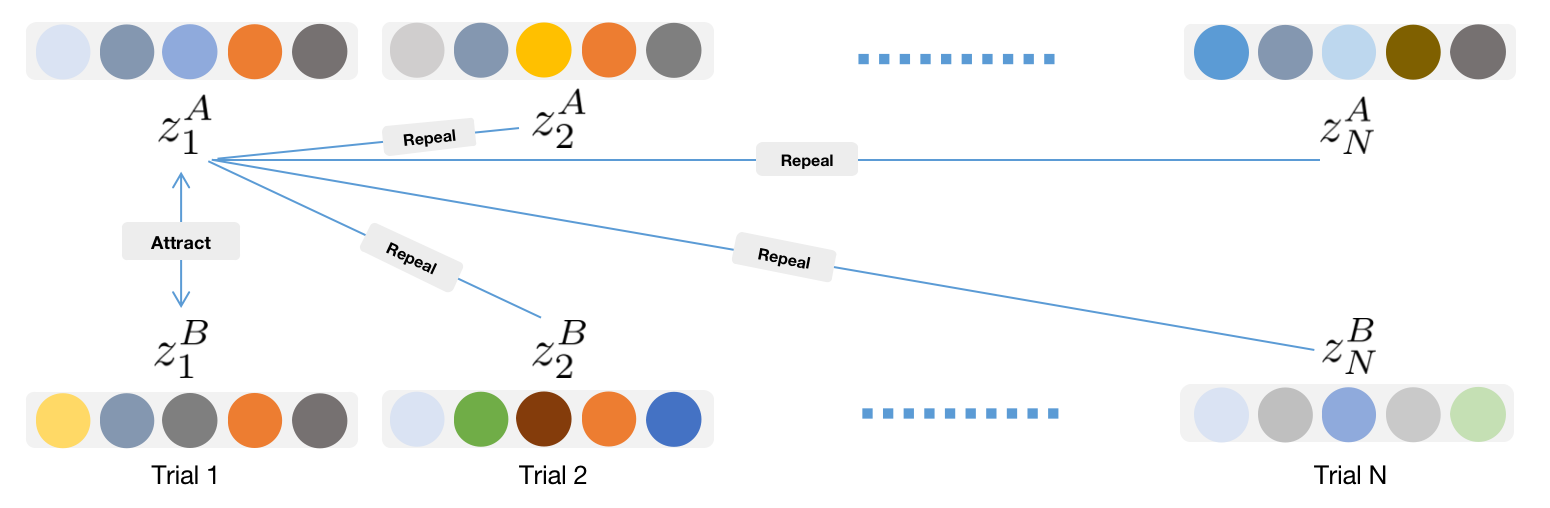}
  \caption{The illustration of the construction of positive pair and negative pair in graph contrastive learning. In a minibatch, embedding $z_1^A$ forms a positive pair with $z_1^B$, other embeddings forms negative pairs with embedding $z_1^A$.}
  \label{fig:contra_illu}
\end{figure}

 

\subsubsection{The functional Graph Contrastive Learning (fGCL) Encoder}
The Graph contrastive learning encoder $f$ takes N-paired graphs (minibatch) $\{\mathcal{G}_i^A|i = 1,..,N \}$ and $\{\mathcal{G}_j^B|j = 1,..,N\}$ as inputs to generate subject-invariant embeddings $\{z_i^A|i = 1,..,N \}$ and $\{z_j^B|j = 1,..,N \}$ of graphs for each type of trial (correct or wrong feedback in our experiment). It adopts spectral graph convolution network consisting of 2 Chebshev spectral graph convolution layers \cite{DBLP:journals/corr/DefferrardBV16} each followed by a TopK pooling layer \cite{cangea2018sparse} as model backbone to capture representations and retain important nodes in the iterative aggregation, global average pooling layer and global mean pooling layer are used to capture the global information, the final two-layer multi-layer perceptron (MLP) is used to output the graph embedding. Similar to SimCLR framework \cite{chen2020simple} with InfoNCE loss, we adopt the contrastive loss function for the anchor embedding $z_i^A$ defined as:
\begin{equation}
    \label{eq:conloss}
    L(z_i^A) = -log(\frac{exp(sim(z_i^A,z_i^B)/\tau)}{\sum^{N}_{j=1}I_{j\neq i} exp(sim(z_i^A,z_j^A)/\tau) + \sum^{N}_{j=1}exp(sim(z_i^A,z_j^B)/\tau)})
\end{equation}
where $I_{j\neq i} = \{0, 1\}$ is an indicator, which is set to 1 when $j \neq i$, $\tau$ is a temperature factor to adjust the attractiveness strength. Overall, this loss function increases the attractiveness of positive pair ($z_i^A$ and $z_i^B$) and decrease the attractiveness of negative pair ($z_i^A$ and others) in the embeddding space. The similarity between two embeddings is computed by
\begin{equation}
    \label{eq:sim}
    sim(z_i^A,z_j^B) = \frac{z_i^A \cdot z_j^B}{\lvert \lvert z_i^A \rvert \rvert \lvert \lvert z_j^B \rvert \rvert}
\end{equation}

Eventually, the accumulated loss of the minibatch is computed by
\begin{equation}
    \label{eq:consum}
    L = \sum_{i=1}^{N} L(z_i^A) + \sum_{i=1}^{N} L(z_i^B)
\end{equation}




The spectral convolution blocks of fGCL comprise of Chebyshev spectral graph convolutional operator (i.e., ChebConv layer), which is defined by:

\begin{equation}
X' = \sum_{k=1}^{K} \mathcal{Z}^{(k)}(X) \cdot \mathcal{\theta}^{(k)}
\end{equation}
where K is the Chebyshev filter size, $\mathcal{Z}^{(k)}(X)$ is computed recursively by $\mathcal{Z}^{(1)} = X$, $\mathcal{Z}^{(2)} = \Tilde{L} \cdot X$, all the way to $\mathcal{Z}^{(k)} = 2 \cdot \Tilde{L} \cdot\mathcal{Z}^{(k-1)} - \mathcal{Z}^{(k-2)}$, 
$\Tilde{L}$ denotes the scaled and normalized Laplacian $\frac{2L}{\lambda_{max}} - I$, where $\lambda_{max}$ is the largest eigenvalue of $L$. $\mathcal{\theta}^{(k)}$ are learnable parameters. To prevent overfitting, each ChebConv layer is followed by a TopK pooling layer, which downsamples graphs and reduce their dimensionality while retaining the most relevant nodes (i.e., top K nodes)  leading the model to focus on meaningful information. Afterwards, embeddings of graphs optimized by graph contrastive learning are input to the downstream classifier to perform graph classification task.

\subsubsection{Downstream Analysis on the Outcome of Emotional Contagion}
After graph contrastive learning phase, the subject difference has been eliminated in the embedding space, that is, the representation extracted the neural activity pattern of the group commonality. To fully utilize these aligned representations, we adopt Dynamic Graph Classification (DGC) model to perform brain emotional contagion state analysis, the trained classifier could better classify the commonality. Initially, the DGC takes a population graph with isolated nodes as the input, by iteratively projecting node features to a new feature space, it is able to construct new edges based on the top-K connection strengths definded as: 
\begin{equation}
\label{eq:dgc}
v_i^{P\prime} = \sum_{m \in \mathcal{E}_{(i, \cdot)}^{P}} \phi(v_i^P \,||\, v_m^P - v_i^P), \quad \mathcal{E}^{P} = \text{KNN}(\mathcal{V}^P)_{\text{topK}}
\end{equation}

The classification tasks are mainly carried out according to the early, middle and late stages of the problem solving tasks. (1) At the early stage, we hypothesized that there was no significant difference in the brain activity of DMT-PST dyads, that is, the emotional contagion effect did not occur, so the differentiation between DMT and PST was low. (2) At the middle stage, we assume that the DMT in DMT-PST dyad begins to induce emotional contagion effect, gradually transferring the negative emotions caused by pressure to PST, so it can be seen that the differentiation between and will gradually increase. (3) At the late stage, we assume that the emotional contagion effect of the DMT-PST dyad has terminated, and the negative emotion of DMT is transferred to PST. Therefore, there is a significant difference in brain activity between DMT and PST, so the differentiation between them reaches the highest. In general, if the distinguishable degree is described with time as the independent variable, then according to our hypothesis, it can be found that the distinguishable degree in the pair will show a positive correlation trend. For the control group PST-PST dyad, we assume that these is no significant discrepancy between subject exists and thus the distinguishable degree displays a stable trend.

The algorithm of training the fGCL encoder and the downstream classifier DGC are summarised in Algotithm \ref{alg:training}.



\begin{algorithm}
\caption{The Training Algorithm of fGCL+DGC --- Learning Graph Embedding \& Classification}
\label{alg:training}
\begin{algorithmic}[1]
\REQUIRE Collection of N batched double view data $\mathcal{B}$, the learning rate $\alpha_l$ of fGCL, the learning rate $\alpha_g$ of DGC, batch size $N$, temperature $T$, ratio $r$, kernel size $K$, maximum number of training epochs of the fGCL \textit{$MaxEpoch_f$}, maximum number of training epochs of the DGC \textit{$MaxEpoch_d$};
\STATE Initialize the parameters of ChebConv and TopKPooling blocks in fGCL with $K$ and $r$ and set $Epochs_f \leftarrow 1$, $Epochs_d \leftarrow 1$;
\WHILE{not converge and $Epochs_f \leq MaxEpoch_f$}
    \FOR{batch $B_b = \{\{(\mathcal{G}_1^1, y_1^1), (\mathcal{G}_2^1, y_2^1)\}, ....,\{\{(\mathcal{G}_1^N, y_1^N), (\mathcal{G}_2^N, y_2^N)\}\}\} \in \mathcal{B}$}
        \STATE Obtain graph embeddings $\{z_i^b|i = 1,2, ..., 2N\}$ by two ChebConv and TopKPooling blocks and a two-layer MLP;\\
        \STATE Calculate graph contrastive loss by (\ref{eq:conloss}) - (\ref{eq:consum});\\
        \STATE $\theta_f \leftarrow \theta_f - \alpha_l \frac{\partial L}{\partial \theta_f}$
        \STATE $\alpha_l \leftarrow F_l(\alpha_l)$ where $F_l(\cdot)$ is the multi step learning rate scheduler;\\
    \ENDFOR
    \STATE $Epochs_f \leftarrow Epochs_f + 1$;
\ENDWHILE

\STATE Construct a population graph $\mathcal{G}_{p}$ on $\mathcal{B}$ with isolated nodes using trained $fGCL_{\theta_f}$;
\WHILE{not converge and $Epochs_d \leq MaxEpoch_d$}
    \STATE Dynamically reconstruct population graph $\mathcal{G}_{p}$ using (\ref{eq:dgc});
    \STATE Calculate focal loss by (\ref{eq:focal}) - (\ref{eq:pt});
    \STATE $\theta_d \leftarrow \theta_d - \alpha_g \frac{\partial L}{\partial \theta_d}$;
    \STATE $Epochs_d \leftarrow Epochs_d + 1$;
    \STATE $\alpha_g \leftarrow F_g(\alpha_g)$ where $F_g(\cdot)$ is the step learning rate scheduler;
\ENDWHILE
\STATE \textbf{Output}: fGCL encoder parameters $\theta_{f}$ and DGC classifier parameters $\theta_{d}$;
\end{algorithmic}
\end{algorithm}

\section{Experiment}



\subsection{Participants}
Eighteen white female students who granted written consent participated in the study for payment. Participants were recruited for the study if they expressed knowledge of the stereotype that men are better at math than women. Specifically, all participants responded with a three or lower to the following question during a pre-study screening: "Regardless of what you think, what is the stereotype that people have about women's and men's math ability" (1= Men are better than women; 7= Women are better than men). Participants were paired into nine dyads. One participant was excluded from EEG analyses due to a lack of valid trials. Thus one (PST) dyad was removed from all dyadic analyses.


\subsection{Procedure}
Upon arrival to the lab, partners of each dyad met for the first time while signing consent forms; they were then prepared for EEG recording. Each member of the dyad was seated in their own soundproof chamber in front of a computer screen and iPad tablet. Dyads were randomly assigned to either an SBS/diagnostic math test condition (DMT, n=14 dyads) or a control/problem-solving task condition (PST, n=15 dyads). In the DMT condition, one participant (referred to as the “actor”) was exposed to SBS by being told they would complete tasks that were diagnostic of their math intelligence. They also completed demographic questions that included a gender query, had pre-recorded instructions read aloud to them in a male voice through headphones, and were prepped for EEG recording by at least one male experimenter. In contrast, the DMT actor’s interaction partner (referred to as the “partner”) and all participants in the PST condition were informed that they would be completing tasks that would inform researchers about the different types of problem-solving techniques they prefer \cite{Forbes_article, Forbes_article2}, completed demographic questions that excluded the gender query, had prerecorded instructions read aloud to them by a female voice through headphones, and were set up by female experimenters. Thus, DMT partners and both participants in the PST condition were always placed in stereotype neutral/stress-free contexts. That is, only the condition of the actor varied across dyad conditions. After an initial set of instructions, participants were connected via webcam on their iPad tablet in order to facilitate face-to-face communication during the interactive math task (described below). Participants were able to see one another through the duration of the interactive math task. When the interactive math task was completed, participants answered a series of questionnaires alone (iPads were removed from the EEG chambers), were debriefed, and were compensated for their participation with cash or course credit.

\subsection{Interactive Math Task}
Actors and partners simultaneously completed a 100 problem math task consisting of standard multiplication and division problems (e.g., 10x20=) that they solved both alone and together. Initial pilot tests confirmed that the problems selected varied in degree of difficulty (easy, medium, and hard), ensuring all participants would solve problems correctly and incorrectly, thus exposing them to both positive and negative performance feedback. Actors and partners were first presented with the same math problem to solve alone for 16 seconds. During this solo time, participants were given three answer choices below each problem (A, B, or C), with the answer to each problem randomly presented in one of the three answer positions. Participants mentally completed all problems without scratch paper and made all answer selections via a button box placed in their laps. This solo answer was used for all performance outcomes in our analyses. After participants entered their solo answer, they were prompted with a screen that said, “Please discuss the answer to the problem with your partner.” At this time, participants were given 20 seconds to discuss their answer with their partners. Participants were then given five seconds to change or confirm their answer to the math problem they just solved alone. After submitting their final response, participants received feedback for two seconds that indicated whether their final answer was correct or incorrect (presented as the words \textit{CORRECT} or \textit{WRONG} written in black on a white screen).




\subsection{EEG Recording}
Consistent with Forbes's research \cite{Forbes_article3}, continuous EEG activity was recorded from each member of the dyad using an ActiveTwo head cap and the ActiveTwo Biosemi system (BioSemi, Amsterdam, Netherlands). Recordings were collected from 64 Ag-AgCl scalp electrodes and from bilateral mastoids. Two electrodes were placed next to each other 1 cm below the right eye to record eye-blink responses. A ground electrode was established by BioSemi’s common Mode Sense active electrode and Driven Right Leg passive electrode. EEG activity was digitized with ActiView software (BioSemi) and sampled at 2048 Hz. Data was downsampled post-acquisition and analyzed at 512 Hz. In our data-driven analysis, we only consider the feedback trials (i.e., \textit{CORRECT} or \textit{WRONG} responses) to better capture elicited emotional patterns when participants are evoked by stimuli.

\subsection{Data Preprocessing}
For feedback analyses, the EEG signal was epoched and stimulus-locked from 500ms pre-feedback presentation to 2000ms post-feedback presentation. EEG artifacts were removed via FASTER (Fully Automated Statistical Thresholding for EEG artifact Rejection) \cite{Nolan_article}, an automated approach to cleaning EEG data that is based on multiple iterations of independent component and statistical thresholding analyses. Specifically, raw EEG data was initially filtered through a band-pass FIR filter between 0.3 and 55 Hz. The EEG channels with significant unusual variance (absolute z score larger than 3 standard deviations from the average), mean correlations with other channels, and Hurst exponents were removed and interpolated from neighboring electrodes using a spherical spline interpolation function. EEG signals were then epoched, and baseline corrected. Epochs with significant unusual amplitude range, variance, and channel deviation were removed. The remaining epochs were then transformed through ICA. Independent components with significant unusual correlations with EOG channels, spatial kurtosis, slope in the filter band, Hurst exponent, and median gradient were subtracted and the EEG signal was reconstructed using the remaining independent components. Finally, EEG channels within single epochs with significant unusual variance, median gradient, amplitude range, and channel deviation were removed and interpolated from neighboring electrodes within the same epochs.

\subsection{Source Reconstruction}
All a priori sources used in network connectivity analyses were identified and calculated via forward and inverse models utilized by MNE-python \cite{Gramfort_article, bb2c8cb6d0ad4c249ee037290ca3669f}. The forward model solutions for all source locations located on the cortical sheet were computed using a 3-layered boundary element model \cite{Hmlinen1989RealisticCG}, constrained by the default average template of anatomical MNI MRI. Cortical surfaces extracted with FreeSurfer were sub-sampled to approximately 10,240 equally spaced vertices on each hemisphere. The noise covariance matrix for each individual was estimated from the pre-stimulus EEG recordings after preprocessing. The forward solution, noise covariance, and source covariance matrices were used to calculate the dynamic statistical parametric mapping (dSPM) estimated inverse operator \cite{dale1999cortical, dale2000dynamic}. The inverse computation was done using a loose orientation constraint (loose = 0.2, depth = 0.8) \cite{Lin_article}. dSPM inverse operators have been reported to help characterize distortions in cortical and subcortical regions and improve the bias accuracy of neural generators in deeper structures, e.g., the insula \cite{attal:hal-00816043} by using depth weighting and a noise normalization approach. The cortical surface was divided into 68 anatomical regions (i.e., sources) of interest (ROIs; 34 in each hemisphere) based on the Desikan–Killiany atlas \cite{Desikan_article} and signal within a seed voxel of each region was used to calculate the power within sources and phase locking (connectivity) between sources. All graph theory calculation descriptions are located in the supplementary materials.

\subsection{Implementation Details} \label{sec:imp_de}
After the data preprocessing, 18 female individuals formed 9 pairs consisting of 7 DMT-PST dyads and 2 PST-PST dyads. We then applied a sliding window technique to all basic views (the initial multi-ROI EEG time series without augmentation), denoted as $\mathcal{X}$, which has 768 time points and 68 ROIs. The width and step size of the sliding window were set to 300 and 50, respectively, resulting in 10 augmented views (68 * 300) for each basic view (68 * 768). This augmentation increased the size of our dataset to 17,650 response graphs, where subjects received either \textit{WRONG} or \textit{CORRECT} feedback. This augmentation allows the model to learn more general patterns. The dataset was split into training, testing, and validation sets in a 7:2:1 ratio. 

In the graph contrastive learning procedure, within a N-sized minibatch, given a graph collection $\{\mathcal{G}_i^j| j = 1, 2, 3, ..., K\}$ representing i-th mathematical question, where $j$ is an index denoting augmented views, $i = 1, 2, ...., N$. We enumerate all possible pairs within these graphs to form positive pairs. Within a dyad $D_m$ consists of subject $D_m^A$ and $D_m^B$, since K augmented views $\{\mathcal{G}_i^j| j = 1, 2, 3, ..., K\}$ of one of the members (e.g. $D_m^A$) represent the same mathematical question and exhibit similar neural activity patterns. Given one view $\mathcal{G}_i^j$ within these K augmented views, the views $\{\mathcal{G}_i^k| k = 1, 2, 3, ..., K, k \neq j\}$ form positive pair with it and negative pairs with views representing other mathematical questions (i.e., $\{\mathcal{G}_r^n| n = 1, 2, 3, ..., K, r \neq i\}$). Additionally, when two subjects encounter the same mathematical question, one augmented view of subject $D^A_m$ in the m-th dyad should form positive pairs with the other K augmented views of subject $D^B_m$ in the same dyad. This process results in a total of 2K * (2K - 1) / 2 positive pairs, which is 190 in our experiments. We trained the model for 700 epochs with early stopping and adopted a multi step learning rate scheduler and Adam optimizer. The initial learning rate, the weight decay, the temperature and the batch size was empirically set to 0.001, 0.02, 0.5 and 68 respectively. The filter size in the Chebyshev spectral graph convolutional operator was set to 4, and the ratio of the TopK pooling layer was set to 0.5.


After obtaining the encoder $f_{\theta}$, graph features can be extracted as a embedding $z_i = f_{\theta}(\mathcal{G}_i)\in R^d$ in the $d$ dimensional embedding space forming isolated nodes in the population graph initially.   
In the feedback type classification procedure, we use the raw dataset (without sliding window augmentation) to train the DGC classifier for 100 epochs. We adopted focal loss eq. (\ref{eq:focal}) - eq. (\ref{eq:pt}) for the classifier to mitigate the aftermath of imbalanced data. We set $\alpha$ and $\gamma$ to 0.5 and 2 empirically.
\begin{equation}
    \label{eq:focal}
    FL(p_t) = -\alpha_t(1-p_t)^{\gamma}log(p_t)
\end{equation}

\begin{equation}
    \label{eq:pt}
    p_t = \begin{cases}
        p & \text{if } y=1 \\
        1 - p & \text{otherwise},
    \end{cases}
\end{equation}

All methods are implemented with Pytorch and trained with GPU NVIDIA\_GeForce\_RTX\_3090.
\section{Experimental Result}
\subsection{Graph Embedding Effectiveness Analysis}
\begin{figure}[htbp]
    \centering

    \includegraphics[scale=0.6]{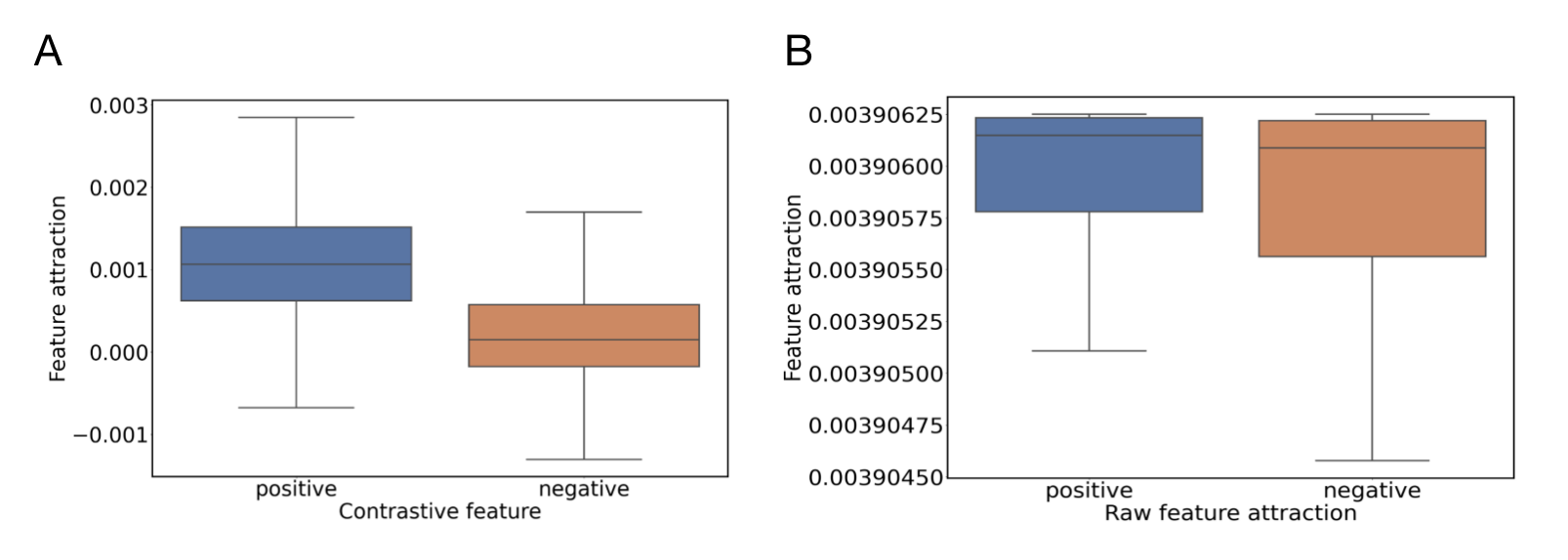}

    \caption{The feature attraction of graph contrastive features and raw features.}
    
    \label{fig:combined}
\end{figure}

\begin{figure}[htbp]
    \centering
        \includegraphics[scale=0.6]{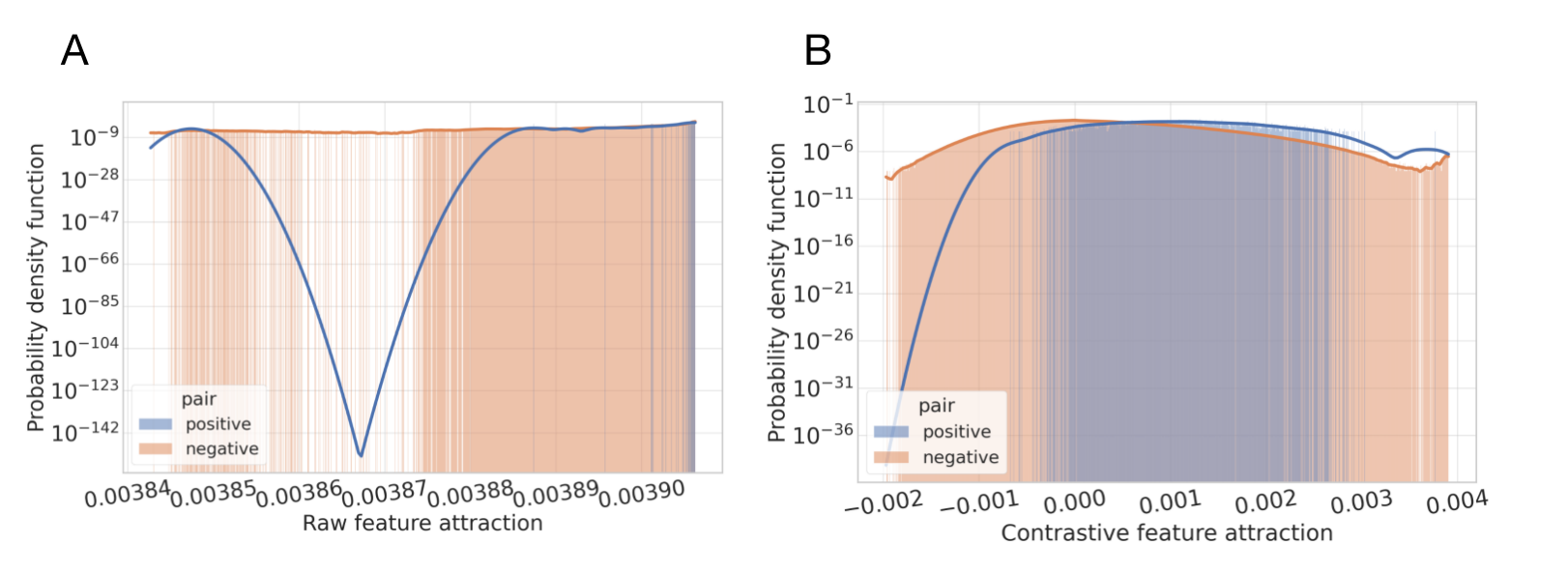}
    \caption{The probability density function of attractions on graph contrastive features and raw features.}
    
    \label{fig:combined2}
\end{figure}

We visualize the feature attraction of graph embeddings and raw features in the dataset. Figure \ref{fig:combined} and Figure \ref{fig:combined2} indicate that the positive pairs have higher feature attraction than that of negative pairs. Besides, the feature attraction of negative pairs is close to 0, which verifies the effectiveness of fGCL in reducing the attraction of negative pairs and heightening the attraction of positive pairs. Conversely, the feature attraction on raw features display no significant differences, indicating identical similarity between positive and negative pairs.


\subsection{Impact of Emotional Contagion on Performance Over Time}
Initial examinations of performance revealed a condition by time interaction, $Wald_{\chi
}^{2}(2)$ = 13.72, p =.001. Specifically, DMT partner’s performance decreased over time (B = -.005 (SE = .002); $Wald_{\chi}^{2}(1)$ = 7.45, (95\% Wald LL CI = -.009; UL CI = -.002), p =.006; for every one unit increase in time (trial number), the log odds of getting the question correct decreased by .005 units.
In contrast, DMT actors and PST dyad members exhibited no over-time changes. Their performance remained stable (p's > .11). Moreover, simple contrasts between conditions revealed performance differences between DMT actors and partners over the course of the task. At the beginning of the task, DMT partners had a higher probability of getting a question correct in comparison to the DMT actor (p=.002), and PST dyad members (p=.034). At the end of the task, DMT partners had a lower probability of getting a question correct in comparison to the partner (p=.004) and were largely comparable to PST dyad members (p=.089). DMT actors did not differ from PST dyad members (p=.35). Thus, at the end of the task, DMT partners underperformed in comparison to the actors. These findings support the possibility that DMT actors benefited from dyadic interactions at the expense of their partners.

\subsection{Contagion Stage Analysis}
\label{se:Contagion_Stage_Analysis}
In the analysis of the contagion stage, we divided the basic view dataset into training, testing, and validation sets in a 7:2:1 proportion. Subsequently, we extracted subject-invariant embeddings using the trained graph contrastive encoder for the classification task.

Table \ref{tab:table_ablation} presents the results of ablation experiments conducted on various models, including KNN, SVM, MLP, and DGC, aimed at classifying feedback types within the basic view dataset. Notably, we employed embeddings extracted by the fGCL encoder as inputs for the KNN, SVM, and MLP models. The experimental results demonstrate that our model exhibits superior performance in the task of feedback type classification while maintaining a balance between sensitivity and specificity. However, it is worth noting that the absence of the fGCL encoder leads to a noticeable degradation in the performance of the DGC model.

\begin{table}[!htbp]
    \centering
    \scalebox{0.9}{
        \begin{tabular}{lccccc}
        \hline
            \textbf{Models} & \textbf{ACC} & \textbf{AUC} & \textbf{F1 Score}  & \textbf{SEN} & \textbf{SPEC}   \\
        \hline
            \textit{fGCL + KNN} & 63.80 $\pm$ 0.15\% & 66.07 $\pm$ 0.55\% & 62.93 $\pm$ 0.15\% & 58.63 $\pm$ 0.43\% & 63.96 $\pm$ 0.05\% \\ 
            \textit{fGCL + SVM} & 59.11 $\pm$ 0.22\% & 63.19 $\pm$ 0.39\% & 52.32 $\pm$ 0.25\% & 51.21 $\pm$ 0.53\% & 55.96 $\pm$ 0.45\% \\ 
            \textit{fGCL + MLP} & 58.11 $\pm$ 0.32\% & 63.19 $\pm$ 0.39\% & 58.61 $\pm$ 0.25\% & 54.38 $\pm$ 0.53\% & 58.13 $\pm$ 0.18\% \\ 
            \textit{DGC w/o encoder} & 61.11 $\pm$ 0.12\% & 63.19 $\pm$ 0.39\% & 58.61 $\pm$ 0.25\% & 57.00 $\pm$ 0.53\% & 64.96 $\pm$ 0.45\% \\ 
            \textbf{fGCL + DGC (ours)} & \textbf{65.12 $\pm$ 0.45\%} & \textbf{69.79 $\pm$ 0.35\%} & \textbf{64.90 $\pm$ 0.06\%} & \textbf{64.71 $\pm$ 0.35\%} & \textbf{65.53 $\pm$ 0.05\%} \\
        \hline
        \end{tabular}
    }
    \caption{The whole-brain level feedback type classification result in the entire period}

    \label{tab:table_ablation}
\end{table}


\subsubsection{Individual-level Classification}
We further examined our hypothesis that whether the emotional contagion occur over the period of sustained interpersonal interaction. To this end, we divided the trial sequence into three stages averagely: early stage ($0 \sim 1/3$), middle stage ($1/3 \sim 2/3$) and late stage ($2/3 \sim 1$) and evaluate them separately by our model. The result in Table \ref{tab:table_DMT_PST} revealed that the classification performance of DMT actors in DMT-PST dyad is better than DMT partners in DMT-PST dyad in Table \ref{tab:table_PST_PST} entirely, and the accuracy and the F1 score of DMT actors in DMT-PST dyad represents an evident increasing trend especially from the early stage to the middle stage and the late stage was superior to the other stages. In contrast, that of DMT partners in DMT-PST dyad displayed a relatively less increment in the accuracy and F1-score. This may additionally reflect that the emotional contagion evolve cumulatively over the sustained interpersonal communication, this in turn, affect the neurological representation when encountering negative event, and thus resulting in the poor classification result of recognizing the pattern of neural response of DMT partners in DMT-PST dyad. On the contrary, DMT actors in DMT-PST dyad release negative emotions by transmitting them to their partners, this then gradually break the disordered response pattern resulting in the improvement of classification performance.

\begin{table}[!htbp]
    \centering
    \begin{tabular}{lccccc}
    \hline
       \textbf{Periods} & \textbf{ACC} & \textbf{AUC} & \textbf{F1 Score}  & \textbf{SEN} & \textbf{SPEC}   \\
    \hline
        \textit{Entire} & 68.28 $\pm$ 0.05\% & 70.49 $\pm$ 0.05\% & 70.50 $\pm$ 0.06\% & 66.67 $\pm$ 0.08\% & 67.63 $\pm$ 0.20\% \\ 
        \textit{Early} & 60.66 $\pm$ 0.14\% & 64.13 $\pm$ 0.06\% & 64.21 $\pm$ 0.17\% & 58.85 $\pm$ 0.29\% & 65.62 $\pm$ 0.25\% \\ 
        \textit{Middle} & 69.33 $\pm$ 0.25\% & 64.35 $\pm$ 0.27\% & 68.23 $\pm$ 0.31\% & 64.51 $\pm$ 0.35\% & 64.52 $\pm$ 0.34\% \\ 
        \textit{Late} & 71.33 $\pm$ 0.17\% & 73.31 $\pm$ 0.06\% & 72.64 $\pm$ 0.22\% & 72.83 $\pm$ 0.28\% & 67.50 $\pm$ 0.31\% \\
        
    \hline
    \end{tabular}
    \caption{The classification result of DMT in DMT-PST dyad at early, middle and late stage with entire testing.}
    
    \label{tab:table_DMT_PST}
\end{table}

\begin{table}[!htbp]
    \centering
    \begin{tabular}{lccccc}
    \hline
        \textbf{Periods} & \textbf{ACC} & \textbf{AUC} & \textbf{F1 Score}  & \textbf{SEN} & \textbf{SPEC}   \\
    \hline
        \textit{Entire} & 63.33 $\pm$ 0.05\% & 67.87 $\pm$ 0.06\% & 67.20 $\pm$ 0.08\% & 66.67 $\pm$ 0.10\% & 63.01 $\pm$ 0.13\% \\ 
        \textit{Early} & 59.38 $\pm$ 0.25\% & 62.12 $\pm$ 0.13\% & 64.96 $\pm$ 0.35\% & 60.00 $\pm$ 0.46\% & 67.14 $\pm$ 1.17\% \\ 
        \textit{Middle} & 64.21 $\pm$ 0.15\% & 63.19 $\pm$ 01.39\% & 68.59 $\pm$ 0.26\% & 68.20$\pm$ 0.37\% & 64.96 $\pm$ 01.45\% \\ 
        \textit{Late} & 64.76 $\pm$ 0.12\% & 68.16 $\pm$ 0.08\% & 65.18 $\pm$ 0.14\% & 67.70 $\pm$ 0.18\% & 65.12 $\pm$ 0.21\% \\
        
    \hline
    \end{tabular}
    \caption{The classification result of PST in DMT-PST dyad at early, middle and late stage with entire period testing.}
    
    \label{tab:table_PST_PST}
\end{table}

Our classifier was evaluated using leave-dyad-out cross validation, results of DMT-PST dyad and PST-PST dyad are shown in Table \ref{tab:table_DMT_PST} and Table \ref{tab:table_PST_PST} respectively, which shed light on our aforementioned hypothesis. Regarding the result of DMT actors inside DMT-PST dyads, our model achieved the classification accuracy as 60.66 $\pm$ 0.14\% at the early stage, this accuracy than increased at the middle stage of problem solving to 69.33 $\pm$ 0.25\% than raised to 71.33 $\pm$ 0.17\% at the late stage. Conversely, the results for DMT partners within DMT-PST dyads displayed a distinct pattern. The accuracy increased from 59.38 $\pm$ 0.25\% in the early stage to 64.21 $\pm$ 0.15\% in the middle stage, with marginal change observed in the late stage, yielding an accuracy of 64.76 $\pm$ 0.12\%. Analyzing this pattern, it becomes apparent that in the early stage of problem-solving, no substantial difference in brain activity existed between DMT and PST pairs, indicating that the emotional contagion effect was relatively dormant. As the problem-solving process advanced to the middle stage, the emotional contagion effect commenced within DMT-PST pairs, with DMT actors gradually transmitting the negative emotions stemming from pressure to DMT partners. As a result, the differentiation between the two groups demonstrated an increasing trend. In the final stage of problem-solving, the emotional contagion effect within DMT-PST pairs subsided, with DMT actors transferring their negative emotions to DMT partners. Consequently, the discernible distinction in brain activity between DMT actors and DMT partners became significant, accentuating the differentiation in neural pattern between the two. Besides, the results of PST-PST dyads in Table \ref{tab:table_PST1} and Table \ref{tab:table_PST2} display no significant changes in the accuracy, this might suggest that the emotional contagion does not appear in PST-PST dyads and thus the neural pattern differentiation in all stages is vague. 

\begin{table}[!htbp]
    \centering
    \begin{tabular}{lccccc}
    \hline
        \textbf{Periods} & \textbf{ACC} & \textbf{AUC} & \textbf{F1 Score}  & \textbf{SEN} & \textbf{SPEC}   \\
    \hline
        \textit{Entire} & 75.03 $\pm$ 0.01\% & 85.54 $\pm$ 0.09\% & 82.48 $\pm$ 0.01\% & 72.06 $\pm$ 0.01\% & 68.09 $\pm$ 0.05\% \\ 
        \textit{Early} & 76.67 $\pm$ 0.05\% & 80.62 $\pm$ 0.25\% & 79.17 $\pm$ 0.15\% & 67.31 $\pm$ 0.03\% & 63.96 $\pm$ 0.05\% \\ 
        \textit{Middle} & 75.11 $\pm$ 03.12\% & 84.19 $\pm$ 01.39\% & 84.35 $\pm$ 0.13\% &  75.61 $\pm$ 0.25\% & 74.96 $\pm$ 0.45\% \\ 
        \textit{Late} & 74.12 $\pm$ 01.45\% & 83.79 $\pm$ 0.35\% & 83.21 $\pm$ 0.11\% & 64.71 $\pm$ 0.35\% & 63.53 $\pm$ 0.05\% \\
        
    \hline
    \end{tabular}
    \caption{The classification result of PST1 in PST-PST dyad at early, middle and late stage with entire period testing.}
    
    \label{tab:table_PST1}
\end{table}

\begin{table}[!htbp]
    \centering
    \begin{tabular}{lccccc}
    \hline
        \textbf{Periods} & \textbf{ACC} & \textbf{AUC} & \textbf{F1 Score}  & \textbf{SEN} & \textbf{SPEC}   \\
    \hline
        \textit{Entire} & 59.83 $\pm$ 0.03\% & 66.38 $\pm$ 0.03\% & 69.13 $\pm$ 0.02\% & 58.46 $\pm$ 0.05\% & 63.96 $\pm$ 0.05\% \\ 
        \textit{Early} & 57.17 $\pm$ 0.31\% & 61.69 $\pm$ 0.21\% & 63.37 $\pm$ 0.22\% & 56.19 $\pm$ 0.43\% & 62.96 $\pm$ 0.05\% \\ 
        \textit{Middle} & 56.31 $\pm$ 0.19\% & 60.12 $\pm$ 0.14\% & 59.37 $\pm$ 0.13\% & 53.19 $\pm$ 0.21\% & 55.36 $\pm$ 0.12\% \\ 
        \textit{Late} & 55.97 $\pm$ 0.13\% & 59.12 $\pm$ 0.13\% & 58.13 $\pm$ 0.12\% & 55.39 $\pm$ 0.13\% & 51.96 $\pm$ 0.03\% \\ 
        
    \hline
    \end{tabular}
    \caption{The classification result of PST2 in PST-PST dyad at early, middle and late stage with entire testing.}
    
    \label{tab:table_PST2}
\end{table}

Overall, the findings put forth the notion that DMT actors and those within the PST dyad condition were processing performance feedback in a comparable fashion, particularly when considering the overall span of the task. This alignment is noteworthy, given that DMT actors underwent initial exposure to SBS. Conversely, discernible distinctions emerge in the processing patterns of DMT partners, especially during the middle and later phases. These outcomes lend credence to the idea that SBS contagion unfolds progressively over time. Notably, DMT partners appears to play a role in buffering the negative emotions of DMT actors.


\section{Discussion}
Overall, our findings suggest SBS-based emotional contagion can occur within female dyads in problem-solving contexts and has different consequences on performance for each member of the dyad. While working together on a math task, DMT partners performed worse over time, whereas DMT actors (i.e., those under SBS), performed better and comparable to dyads working in SBS neutral contexts. Importantly, DMT partners showed evidence of “catching” this initial stress response from the threatened actor. This, in turn, had direct ramifications for DMT partners’ performance. Contrastively, these relationships were not evident within PST dyad members interacting in SBS neutral contexts.

\subsection{Inference of performance decrement in dyadic interaction}
Results provide further insight into the dynamic relationship between two individuals performing in domains where their common identity is devalued. Although it seems conceivable that the performance of both DMT actors and partners would suffer when solving problems together in a negatively stereotyped domain, results provide further support for the notion that non-threatened partners help buffer initially threatened actors from the deleterious consequences of SBS over time, at their own expense. These findings are consistent with past work showing that the presence of a female role model or competent female partner alleviates performance decrements otherwise typically evident in stereotype threatening contexts \cite{Marx_article, McIntyre_article, Thorson_article}. Results expand upon this work in several ways. Most notably, by demonstrating that the transference of an individual's stress response on to their partners may be one important factor in buffering women from SBS during dyadic problem-solving interactions, particularly during initial stages of the interaction. Conversely, like past work demonstrating that increased emotional processing of feedback in SBS contexts has a negative impact on individuals’ performance when alone \cite{Forbes_article2, Forbes_article3}, findings from this study demonstrate that this effect extends to partners in a dyadic interaction, providing a potential mechanism for underperformance effects among these individuals in group problem-solving contexts moving forward.

\subsection{Efficiency in timescale and Implications for Contagion Hypotheses}
Present results provide novel insight into how contagion manifests on the order of milliseconds, a much more rapid timescale than previously assumed, to affect performance accordingly using neuroscience methodologies. Moreover, the design of the present study provides a novel yet realistic platform to examine emotion contagion phenomena via EEG or fMRI methodology in future studies. By using iPads, it was possible to capture simultaneous EEG activity in a controlled manner while still allowing participants to have real time face to face interaction. This design also provides implications for contagion hypotheses specific to the mimicry and proximity literature. Because participants only communicated through an iPad webcam, participants were only able to view their partner's face and hear their voice through the webcam during the interaction. This suggests that vocal patterns and facial expressions may have played an integral role in facilitating contagion effects \cite{hatfield1993emotional, neumann2000mood}. Findings provide a more nuanced understanding of the contagion process while also providing a better understanding of a heretofore largely unexamined question in the literature: how social identity threats and SBS manifest in dyadic interactions to have paradoxical effects on performance. More importantly, findings provide further insight into the many ways the gender gap in STEM domains can be perpetuated but also one day nullified.

\subsection{Graph-based Approach for Subject Invariant Emotional Representation Extraction}
Graph structures naturally align with the brain's topology, allowing for the effective modeling of anatomical regions of interest while preserving functional connectivity (FC) information. The proposed fGCL model harnesses semantically meaningful information i.e., graphs corresponding to the same mathematical problem, to construct both positive and negative pairs. This innovative approach significantly mitigates inter-subject variability in EEG data and has been evaluated as optimal. The subject-invariant representations extracted from FC graphs are well-aligned within the common embedding space. Furthermore, the fGCL employs a spectral graph network capable of convolving across the entire node set with FC connections, integrating valuable cognitive information \cite{COHEN2018515}.

\subsection{Limitations and Future Works}
Regarding the limitations of the study, it is important to note that when considering EEG data and the spatial limitations associated with the methodology, conclusions based on precise brain locations should always be interpreted with caution. Results should be replicated and expanded upon in future fMRI studies, although given the temporal constraints of fMRI methodologies with respect to findings in this study (i.e., these effects may occur on the order of milliseconds), this approach could be problematic as well. Besides, the proposed fGCL encoder was validated with EEG data of young female students (mean age = 25.18 years) and optimized with semantical auxiliary task. As is well known that age plays an important role in emotion processing and thus might have different emotion patterns in different age ranges \cite{Ebner_article}, further studies should be conducted to cover different age ranges for a more generalized encoder model. 


\section{Conclusion}
In this study, we addressed a critical issue in previous emotional contagion research: the neglect of subject-level difference. We hereby adopt a self-supervised learning approach to eliminate the subject difference by increasing the attraction of positive pairs and reducing the attraction of negative pairs, which aligns the neurophysiologically meaningful representations of EEG signals in the embedding space, this results in the removal of subject variant information. Based on this substrate, we than employ a dynamic graph classification model to analyze process of emotional contagion. The results suggest that DMT actors and those in PST dyads processed feedback similarly, but distinctions emerged in DMT partners' processing patterns, supporting the idea of gradual SBS contagion. DMT partners appeared to buffer negative emotions of DMT actors.

\section*{Conflict of Interest Statement}
The authors declare that the research was conducted in the absence of any commercial or financial relationships that could be construed as a potential conflict of interest.

\section*{Author Contributions}
\textbf{J.H.: }Conceptualization, Methodology, Formal Analysis, Software, Validation, Writing - Original Draft, Writing - Review \& Editing, Visualization; \textbf{R. A.: }Data Collection; \textbf{M.L.: } Conceptualization, Writing - Review \& Editing; \textbf{C.F.: }Funding Collection. 


\section*{Acknowledgments}
All aspects of this study and article were supported by National Science Foundation grant \#1535414 awarded to Chad E. Forbes.

\section*{Supplemental Data}
None

\section*{Data Availability Statement}
Our native EEG data will be made available on request from any qualified investigator who provides a practicable proposal, or for the purpose of replicating procedures and results presented in the current study. 

\newpage

\bibliographystyle{unsrt}  
\bibliography{references}

\end{document}